\documentstyle[12pt]{article}
\makeatletter
\@addtoreset{equation}{section}
\makeatother
\renewcommand{\theequation}{\thesection.\arabic{equation}}
\newcommand{\bibi}{\bibitem}

\newcommand{\eq}{\ref}
\newcommand{\beq}{\begin{equation}}
\newcommand{\eeq}{\end{equation}}
\newcommand{\bea}{\begin{eqnarray}}
\newcommand{\eea}{\end{eqnarray}}

\newcommand{\cc}{\cite}
\newcommand{\lb}{\label}

\newcommand{\gsim}{\stackrel{>}{\sim}}  

\newcommand{\ssl}{s\!\!\!/}

\newcommand{\half}{\frac{1}{2}}

\newcommand{\vev}{\langle \Phi \rangle}

\newcommand{\un}{1\!\!1}

\newcommand{\bt}{\beta}
\newcommand{\lag}{\langle}
\newcommand{\rag}{\rangle}
\newcommand{\gm}{\gamma}
\newcommand{\pl}{\partial}

\newcommand{\vv}{\langle \Phi \rangle}
\newcommand{\vvs}{\langle \Phi^{st} \rangle}

\newcommand{\kp}{\kappa}

\newcommand{\sg}{\sigma}

\newcommand{\Ps}{\Psi}
\newcommand{\Ph}{\Phi}
\newcommand{\Om}{\Omega}
\newcommand{\Psb}{\overline{\Ps}}

\newcommand{\yb}{\overline{y}}
\newcommand{\hmu}{\hat{\mu}}

\newcommand{\ra}{\rightarrow}

\newcommand{\be}{\begin{equation}}
\newcommand{\ee}{\end{equation}}

\newcommand{\k}{$\kappa$}
\newcommand{\hw}{\widehat{a p}}

\newcommand{\htr}{\; {\textstyle \frac{1}{2} \mbox{Tr}\,}}
\def\lag{\langle}
\def\rag{\rangle}
\def\dateandnumber(#1)#2#3#4{
\vbox to 18mm{%
     \hbox to \textwidth{ \hspace*{14mm} \hsize=40mm%
            \vbox{%
                 \hbox to 40mm{\large #1 \hss}%
                 \hbox to 40mm{    \hss}%
                 \hbox to 40mm{    \hss}%
                 }%
                 \hss \hsize=80mm%
            \vbox{%
                 \hbox to 80mm{\hss \large #2}
                 \hbox to 80mm{\hss \large #3}
                 \hbox to 80mm{\hss \large #4}
                 }%
            \hspace*{14mm} }%
      \vss
    }
}
\def\titleofpreprint#1#2#3#4{{\LARGE \bf
\vbox to 43mm{%
     \vss
     \hbox to \textwidth{ \hspace*{14mm} \hsize=130mm%
            \hss \vbox{
                      \hbox to 130mm{\hss \LARGE \bf #1\hss}%
                      \hbox to 130mm{\hss \LARGE \bf #2\hss}%
                      \hbox to 130mm{\hss \LARGE \bf #3\hss}%
                      \hbox to 130mm{\hss \LARGE \bf #4\hss}%
                 }%
            \hss \hspace*{14mm} }%
      \vss
    }}
}
\def\listofauthors#1#2#3{{\large
\vbox to 22mm{%
     \vss
     \hbox to \textwidth{ \hspace*{14mm} \hsize=130mm%
            \hss \vbox{
                      \hbox to 130mm{\hss \large #1\hss}%
                      \hbox to 130mm{\hss \large #2\hss}%
                      \hbox to 130mm{\hss \large #3\hss}%
                 }%
            \hss \hspace*{14mm} }%
      \vss
    }}
}
\def\listofaddresses#1#2#3{{\small
\vbox to 18mm{%
     \vss
     \hbox to \textwidth{ \hspace*{14mm} \hsize=130mm%
            \hss \vbox{
                      \hbox to 130mm{\hss \small #1\hss}%
                      \hbox to 130mm{\hss \small #2\hss}%
                      \hbox to 130mm{\hss \small #3\hss}%
                 }%
            \hss \hspace*{14mm} }%
      \vss
    }}
}
\def\abstractofpreprint#1{{\normalsize
\vbox to 110mm{%
     \vss
     \hbox to \textwidth{\hss \normalsize \bf Abstract \hss}%
     \vspace*{1cm} \normalsize
     #1
     \vss
    }}
}
\def\footnoteoftitle#1{{\small
\vbox to 30mm{\parindent0pt
     \vss\small #1 \vss
    }}
}
\def\footnoteitem(#1)#2{
\begin{list}{#1}{\labelwidth4.0mm \leftmargin7.0mm
\labelsep2.5mm \rightmargin7.0mm \parsep0.5ex plus0.2ex minus0.1ex
\itemsep0ex plus0.2ex }
\item #2
\end{list}
}
\begin{document}
\dateandnumber(December 1991)%
{HLRZ J\"ulich 91-82}%
{Amsterdam ITFA 91-33}%
{                    }%
\titleofpreprint%
{          Spontaneous symmetry breaking                       }%
{                 on the lattice                               }%
{        generated by Yukawa interaction$^*$                   }%
{                                                              }%
\listofauthors%
{        Wolfgang~Bock$^{**}$, Asit~K.~De, Christoph~Frick,    }%
{                Jiri~Jers\'ak and Thomas~Trappenberg          }%
{                                                              }%
\listofaddresses%
{\em Institute of Theoretical Physics E, RWTH Aachen,           %
     D-5100 Aachen, Germany                                    }%
{                            and                               }%
{\em HLRZ c/o KFA J\"ulich,                                     %
     P.O.Box 1913, D-5170 J\"ulich, Germany                    }%
\abstractofpreprint{
 We study by numerical simulation a lattice Yukawa model
 with naive fermions at intermediate values
 of the Yukawa coupling constant $y$ when the nearest neighbour coupling
 $\kp$ of the scalar field $\Phi$ is very weakly ferromagnetic
 ($\kp \approx 0$) or even antiferromagnetic ($\kp < 0$)
 and the nonvanishing value of $\vev$
 is generated by the Yukawa interaction.
 The renormalized Yukawa coupling
 $y_R$ achieves here its maximal value and
 this $y$-region is thus of particular importance for lattice
 investigations of strong Yukawa interaction.
 However, here the scalar field propagators have a very complex structure
 caused by fermion loop corrections and
 by the proximity of phases with antiferromagnetic properties.
 We develop methods for analyzing these propagators and for extracting
 the physical observables. We find that
 going into the negative $\kp$ region,
 the scalar field renormalization constant
 becomes small and $y_R$ does not seem to exceed the unitarity bound,
 making the existence of a nontrivial fixed point in the
 investigated Yukawa model quite unlikely.
}
\footnoteoftitle{
\footnoteitem($*$){
Supported by the
Deutsches Bundesministerium f\"ur Forschung und Technologie and the
Deutsche Forschungsgemeinschaft
}
\footnoteitem($**$){%
Present address: Institute of Theoretical Physics,
University of Amsterdam, Valckenier\-straat~65,
XE 1018 Amsterdam, The
Netherlands
}
}
\pagebreak

\section{Introduction}

The scalar field self-coupling and
the Yukawa couplings in the electroweak
theory are believed to vanish in the limit of infinite cut-off, as
suggested by the signs of the perturbative $\bt$-functions.
But the confirmation of this ``triviality'' of these couplings and a
reliable determination of its consequences require nonperturbative
methods, because one has to control a very difficult
regime -- when the renormalized couplings have maximal
possible values which could, in principle, be large.
A use of nonperturbative
lattice methods for these purposes is desirable.

In the pure scalar sector of electroweak interactions
these methods have been successful
in estimating the upper bound for the Higgs boson mass in the
approximation neglecting the gauge and fermion fields
\cc{DaNe83}--\cc{LuWe}.
They also indicate that the renormalized quartic coupling
is lower than the upper bound obtained from the tree level unitarity.
This result makes a strongly interacting Higgs sector rather improbable
and explains the quantitative agreement with
calculations using the renormalized perturbation theory.
Weakly coupled gauge fields do not seem to have any unexpected effects
on the $\Phi^4$ theory \cc{HaHa86b} and can thus be treated
perturbatively.

Possible effects of a strong Yukawa coupling remain
relatively unexplored, however.
The most important phenomenological questions are:
\begin{itemize}
\item What would be the effect of a strong Yukawa coupling on the upper
      bound for the Higgs boson mass \cc{Li86}?
\item What is the lower bound on the Higgs boson
      mass which follows
      from the vacuum stability requirement
      in the case of a strong Yukawa coupling
      \cc{Li86,LOWERB}?
\item How strong can the Yukawa coupling actually be?
      What is the maximal fermion mass which can be
      generated by a Yukawa interaction \cc{EiGo86}?
\end{itemize}
{}From the general point of view of quantum field theory the
investigations of models with strong Yukawa coupling attempt to
elucidate the following issues:
\begin{itemize}
\item Are the Yukawa theories in 4 dimensions really trivial
      or do some nontrivial fixed points exist?
\item If they are trivial, how far and in what form Yukawa theories
      can be used as effective quantum field theories?
\item What are their relations to the purely fermionic theories
      with four-fermion coupling \cc{Na88}--\cc{Zi91}?
\end{itemize}

Recently numerous explorations of various lattice Yukawa
models have been performed
\cc{YCONT}--\cc{MIRRORS}
(for a recent review see \cc{DeJe92}).
In the following discussion, only the models with the so-called
lattice parametrization and single-site Yukawa coupling
are considered.
The studies
reveal the existence of three qualitatively different regions of the
bare Yukawa coupling $y$.
If $y$ is sufficiently small, the lattice Yukawa models behave
according to the perturbation theory based on the quasi-classical picture
of the spontaneous symmetry breaking (SSB) in the scalar sector.
For large $y$ the fermions decouple in the continuum limit
so that the physically interesting range of the values of $y$ is
restricted from above.

In the intermediate region, the phase with nonvanishing scalar field
expectation value $\vev$ exists even if the nearest neighbour coupling
$\kp$ of the scalar field is weak or even antiferromagnetic ($\kp < 0$).
The corresponding SSB is then generated by the Yukawa coupling.
In particular, at $\kp=0$ the Yukawa models on the lattice correspond to
the pure fermion theories with multifermion couplings of the
Nambu--Jona-Lasinio type
\cc{Na88}--\cc{Zi91}
and one realizes that
the SSB in such theories can be understood on the lattice as
a special case of the SSB caused by the Yukawa interaction.

In the light of the above-mentioned goals a very important observation
\cc{BoDe91a} is that it is this intermediate
region where
the renormalized Yukawa coupling achieves its maximal values.
Thus there is ample motivation to investigate SSB in Yukawa
theories on the lattice in the regime where the
Yukawa interaction is its driving mechanism and the
standard quasi-classical
approximation based on the scalar field potential is not adequate.

Our recent numerical investigations of the SU(2)$_L\otimes$SU(2)$_R$
lattice Yukawa model with naive lattice fermions have shown
\cc{Ja91,BoDe91a} that in the region of intermediate values of $y$
and negative $\kp$ the scalar field propagators have a very complex
form.
As it turns out, it has two reasons:
Firstly, the Yukawa coupling causes here
sizeable fermion loop corrections to the scalar propagator.
These effects can actually be used to estimate the value of the
renormalized Yukawa coupling.
Secondly, the antiferromagnetic ordering tendency of the negative scalar
field coupling $\kp$
competes with the ferromagnetic
ordering effect of the Yukawa interaction.
This shows up at large momenta
and is therefore a lattice artefact which, however,
has to be taken into account particularly on small
lattices in the analysis of the scalar propagators.

In this paper we
develop reliable methods of analysis of numerical data
in Yukawa models and
present some results for the renormalized Yukawa and
scalar quartic couplings.
The main
improvements over previous investigations of a similar type are
threefold:
\begin{itemize}
\item[(i)] We have been
able to analyze the scalar propagators in a very satisfactory way by
including the 1-fermion loop contribution and as a result
the Goldstone wave function
renormalization constant is now determined quite precisely.

\item[(ii)] To gain experience with finite-size effects
we have varied the lattice
size upto $12^3 16$ and have found the main effect to be the Goldstone
boson dictated $1/L^2$ dependence.

\item[(iii)] We have also covered nearly
the full range of bare parameters
which includes the maximum possible bare Yukawa coupling, staying within
the lattice artefact-free region of the coupling parameter space.
\end{itemize}
However, the fermion number in our simple model
with naive fermions is too large.
This probably causes the most serious
problem we have found, namely that the fermion mass stays
much below the scalar mass even for the
largest renormalized Yukawa coupling.
Therefore phenomenologically relevant large scale
investigations of Yukawa theories should be performed
in more realistic models.

After defining the model in sect.~2 we describe in sect.~3
the spectrum in various phases and point out the appearance
of the ``staggered'' scalar states to be interpreted as lattice
artefacts occuring on a hypercubic lattice in the vicinity
of the phases with antiferromagnetic ordering.
In sect.~4 we discuss the physical meaning of the SSB occuring
at negative $\kp$.
The complex structure of the scalar propagators and the  method
of their analysis on finite lattices are described in sects.~5--7.
In sect.~8 the fermion and scalar masses are discussed.
Some results for the renormalized Yukawa coupling and
the renormalized scalar field quartic coupling are presented
in sects.~9 and 10, respectively.
We summarize our results and conclude in sect.~11.
Also a brief appendix elucidating the properties of models
with both the ferromagnetic and antiferromagnetic orderings
is included.

\section{The SU(2)$_L\otimes$SU(2)$_R$ model with naive fermions}

Our starting point is a fermion-scalar system in the continuum,
which is defined in euclidean space-time by the action
\bea
S_0 &=& \int \mbox{d}^4x
\left[ \,
\frac{1}{2}  \htr  \left\{
(\pl_{\mu} \Phi_0 )^{\dagger} (\pl^{\mu} \Phi_0) \right\}
+ \frac{m_0^2}{2} \; \htr \left\{ \Phi_0^{\dagger}
\Phi_0 \right\} + \frac{g_0}{4!}  \htr \left\{ \left(
\Phi_0^{\dagger} \Phi_0 \right)^2 \right\} \right.  \nonumber \\
&& \hspace{2cm} + \left. \Psb_0 \gm^{\mu} \pl_{\mu} \Ps_0
+ y_0 \Psb_0
\left( \Phi_0 P_R + \Phi_0^{\dagger} P_L \right) \Ps_0
\vphantom{ \left( \frac{1}{2} \right)^2 }
\right] \; . \lb{LA}
\eea
Here $\Phi_0 $ is a 4-component scalar field with a quartic
self-coupling and $\Ps_0$ is a fermionic SU(2) doublet field
coupled to the scalar field by an Yukawa interaction
with the bare Yukawa coupling parameter $y_0$.
$P_R$ and $P_L$ are the right- and left-handed chiral projectors.
Because both fermions in the doublet couple with the same strength $y_0$
to the scalar field, the action has a global SU(2)$_R$ flavour symmetry.
Together with the global SU(2)$_L$ symmetry, which would turn into a
local symmetry when gauge fields are included, the action is invariant
under the global chiral SU(2)$_L\otimes$SU(2)$_R$ transformations
\bea
\Ps_0 & \ra & (\Om_L P_L + \Om_R P_R) \Ps_0, \lb{PSIT} \\
\Psb_0 & \ra & \Psb_0 (\Om_L^{\dagger} P_R + \Om_R^{\dagger} P_L),
\lb{PSIBT} \\
\Ph_0 & \ra & \Om_L^{} \Ph_0 \, \Om_R^{\dagger} \lb{PHIT},
\eea
where $\Om_{L,R} \in$ SU(2)$_{L,R}$.

We regularize the model by introducing a 4-dimensional
hypercubic lattice with the lattice spacing $a$.
A simple possibility is to keep the {\em continuum parametrization}
and just
to replace the derivatives by the lattice differences in the action
(\eq{LA}).
But for a study of the largest renormalized Yukawa coupling
it turns out to be very important to rescale the fields
\be
\Phi_0 (x)=\sqrt{2 \kp} \; \Phi_x / a \;, \;\;\;\;\;\;\;
\Ps_0 (x)   = \Ps_x / a^{3/2}
                                  \lb{A}
\ee
and reparametrize the coupling parameters
\be
(a m_0)^2=\frac{1-2\lambda}{\kp} - 8\;, \;\;\;\;
g_0   =\frac{6 \lambda}{\kp^2} \;, \;\;\;\;
y_0 = \frac{y}{ \sqrt{2 \kp} }\;,     \lb{BC}
\ee
thus ending up with the model in the {\em lattice parametrization}
defined by the action
\bea
S  &=&-\kp \; \sum_{x \mu} \htr
 \left\{ \Ph^{\dagger}_x \Ph^{}_{x + \hmu}
 + \Ph^{\dagger}_{x+ \hmu} \Ph^{}_x \right\}
 + \sum_{x} \htr
   \left\{ \Phi_x^{\dagger} \Phi_x +\lambda \left( \Phi_x^{\dagger}
  \Phi_x - \un \right)^2 \right\}   \nonumber \\
   & &+  \sum_{x \mu} \frac{1}{2}
\left(\Psb_x \gm_{\mu} \Ps_{x + \hmu} -
 \Psb_{x + \hmu} \gm_{\mu} \Ps_x \right)
    + y \;
\sum_x \Psb_x ( \Ph_x^{} P_R + \Ph_x^{\dagger} P_L) \Ps_x \; .\lb{SS}
\eea
We note that all the fields and the parameters in this
expression are dimensionless.
In the following we set
the bare quartic coupling $\lambda$ of the scalar field to
infinity, which leads to a freezing of the radial mode
of the scalar field, $\htr\Phi^\dagger_x \Phi_x = 1$,
so that $\Phi_x$ can be
represented by an SU(2) matrix.
The hopping parameter of the scalar field $\kp$ and the
bare Yukawa coupling $y$ are left as free parameters.
As usual in statistical mechanics, in the action (\eq{SS})
also the values $\kp < 0$ are admissible, though the relationships
(\eq{A}) and (\eq{BC}) to the continuum parametrization
are not defined for negative $\kp$.

For the sake of simplicity we
are using the naive lattice fermions and the model
actually involves 16 degenerate fermion doublets.
There are some approaches trying to achieve a more realistic
fermion spectrum
preserving the chiral symmetry of the action.
So far none of these approaches have proved satisfactory
(see e.g. \cc{BoDe91b} and for recent reviews \cc{DeJe92,CHIRAL}).
An analytic treatment of the naive model by the $1/N$ expansion
is not feasible in the investigated limit
$\lambda\!=\!\infty$.
In order to use the Hybrid Monte Carlo algorithm in the
numerical simulations we further have to double the number of fermions by
squaring the fermion determinant.
Thus we are actually simulating the model with 32 fermion doublets
expecting that qualitative aspects of Yukawa models
are not changed by a large fermion number.

One of such properties, found in many lattice Yukawa models with
different symmetries and numbers of fermions
\cc{YLAT}--\cc{MIRRORS},
is the occurence of SSB even at negative values of the hopping parameter
$\kp$.
Here we face the somewhat puzzling fact that the relation
(\eq{BC}) between the coupling parameters of the continuum
(\eq{LA}) and lattice parametrizations (\eq{SS}) breaks down
for negative $\kp$.
As long as physical (renormalized)
quantities are considered, the region $\kappa<0$ can have a well
defined physical meaning.
The physically interesting region of the
parameter space seems to be extended in the lattice
parametrization and the use of this parametrization is thus crucial.

This observation motivates definitions of renormalized
quantities which do not use the field $\Phi_0$ nor the transformations
(\eq{A}) and (\eq{BC}) and are thus applicable
also for $\kappa < 0$:
The scalar field in lattice parametrization
is renormalized in the phase with SSB as
\be
 \Phi_{x,R}=\frac{\Phi_x}{\sqrt{Z_{\pi}}}    \; ,
\ee
where $Z_{\pi}$ is the wave function renormalization constant
of the Goldstone components of the $\Phi$-field propagator.
For $\kappa>0$ this is equivalent to the renormalized scalar field in the
continuum parametrization
\be
   \Phi_{0,R}(x)=\frac{\Phi_0 (x)}{\sqrt{Z_{0,\pi}}} \;
\ee
when the dimensions of the fields are taken into account,
\be
 \Phi_{x,R} = a \; \Phi_{0,R}(x) \;, \lb{PhiR}
\ee
and
\be
 Z_{0,\pi} = 2 \kp \, Z_\pi \; .  \lb{relZ}
\ee
The vacuum expectation value
of the renormalized scalar field in lattice units
is obtained from the magnetization $\vv$
\be
 a v_R = \frac{\vv}{\sqrt{Z_{\pi}}}    \lb{vR}
\ee
and $v_R$ is the vacuum expectation value
in physical units.
The renormalized
couplings in the broken phase may be defined as
\be
y_R = \frac{m_F}{v_{R}}
=\frac{(a m_F)}{\vv} \; \sqrt{Z_{\pi}} \; , \lb{TR}
\ee
and
\be
\lambda_R =  \frac{m_{\sigma}^2}{2v_{R}^2}
=  \frac{(a m_{\sigma})^2}{2\vv^2} \; Z_{\pi} \;,\lb{LR}
\ee
where $m_F$ and $m_\sg$ are the fermion mass and the $\sg$ boson mass,
respectively, in physical units.

\section{Spectrum and continuum limit at negative~$\kp$}

The phase diagram of the SU(2)$_L\otimes$SU(2)$_R$ model (\eq{SS})
with naive
fermions shown in fig.~1  \cc{BoDe90a,BoDe91a} consists of several
phases and phase regions with ferromagnetic (FM), paramagnetic (PM),
antiferromagnetic (AM) and ferrimagnetic (FI) ordering of the scalar
field.
In addition to the weak coupling phases
FM(W), PMW and AM(W)\footnote{For simplicity
we use the nomenclature ``phases'' also to denote phase regions
with weak (W) or strong (S) bare Yukawa coupling
and include a bracket in the abbreviation.}
we find at large values of $y$ the strong coupling phases
FM(S), PMS and AM(S) with
nonperturbative behaviour of several fermionic observables.
A similar phase structure was also observed in fermion-scalar models with
different symmetry groups and/or different formulations of fermions on
the lattice \cc{YLAT}--\cc{MIRRORS}
provided the lattice parametrization
and single-site Yukawa coupling is used.
Common features of all these models are
(i) the existence of weak and strong Yukawa coupling phases and phase
regions,
(ii) the continuation, at intermediate values of $y$, of the FM phase
into the negative $\kappa$ region and
(iii) the existence of points where several phase transition lines
meet.
The width of the funnel in the phase diagram filled by the FM and FI
phases becomes smaller if the number of fermions is decreased (compare,
e.g., the phase diagrams in refs.~\cc{BoDe90a,BeHe91}) in accordance
with mean field calculations \cc{StTs}.

The fermion mass \cc{BoDe89a} obeys within the error bars
the tree level
relation $am_F \!=\! y\,\vv$ in the FM(W) phase and is zero in
the PMW phase where $\vv=0$.
It increases when the FM(S)--PMS phase transition is approached from the
FM(S) side, does not feel this phase transition and continues to
increase when $\kappa$ is lowered within the PMS phase.

The scalar spectrum is quite different in different phases.
In the four phases around the
point A, where the PMW, AM(W), FM(W) and FI phases meet,
the spectrum is indicated
in fig.~2.
In the phases PMW and AM(W)
with zero magnetization $\vv$
there exists a scalar $\Phi$ particle quadruplet
of degenerate mass $am_{\Phi}$.
On the other hand, in the phases FM(W) and FI with nonzero
magnetization $\vv$ there exist three Goldstone bosons called $\pi$
and one massive $\sigma$ boson
with masses $am_{\pi}\!=\!0$ and $am_{\sigma}$, respectively.

In the $\kp < 0$  region
the scalar propagators show near the momentum
$(\pi,\, \pi,\, \pi,\, \pi)$
the presence of further scalar states which we call
{\em staggered states} in the following.
This effect shows up feebly already at small positive $\kp$
and as $\kp$ is lowered it results in a gradually increasing
curvature in the scalar propagators at large momenta.
The existence of such states is obvious at $y=0$,
as here the symmetry of the action (\eq{SS}) under the
transformation
\be
     \Phi_x \ra \Phi^{st}_x = (-1)^{x_1+x_2+x_3+x_4} \Phi_x , \;\;\;
           \kp \ra -\kp         \lb{ST}
\ee
implies the presence of a $\Phi^{st}$ quadruplet in the PM phase
as well as of three $\pi^{st}$ and \linebreak
one $\sg^{st}$ states in the AM phase
with nonvanishing staggered magnetization \linebreak
$\vvs = \lag \sum_x (-1)^{x_1+x_2+x_3+x_4} \Phi_x \rag$.
They are visible at low momenta in the 2-point
function of the field $\Phi^{st}_x$.
We denote the corresponding masses $am_{\Phi}^{st}$, $am_{\pi}^{st}$
and $am_{\sigma}^{st}$, respectively.
For $0<y<\infty$ the symmetry (\eq{ST}) is broken explicitly, and
these states can appear on the lattice
simultaneously with the usual $\Phi$ or $\pi$ and $\sg$ bosons,
as indicated in fig.~2.

The question is, which particles will remain in the various
possible continuum limits,
or in the large cut-off limits if the theory is an effective one
(several possibilities are discussed in ref.~\cc{DeJe92}).
We want of course to recover in the continuum limit the fermions
and the usual bosons simultaneously.
This includes the renormalized vacuum expectation value
of the scalar field $av_R = \langle \Phi_R \rangle$, which is
proportional to the gauge boson mass $am_W$.
This is possible only in the FM(W) phase
in the scaling region of the FM(W)-PMW
phase transition.
Here the model is of physical relevance from the point of view
of the electroweak theory.

As $am^{st}_\Phi$ and $am^{st}_\sg$ vanish only on the
critical lines where $\vvs$ is vanishing,
the staggered states could remain in the continuum limit
simultaneously with the fermions and the usual scalars
only if the limit
is taken at the point A of the phase diagram in fig.~1.
The scaling region of this point should therefore
for physical reasons probably be avoided.
The staggered states are lattice artefacts
because they depend substantially on the
lattice geometry.
But they can have drastic effects on finite lattices and thus have
to be taken into consideration in the analysis of the numerical data
for the scalar propagator.

\section{SSB generated by Yukawa coupling and the relation to
four-fermion coupling}

The negative $\kp$ region has no analogue in the continuum
parametrization because the transformation equations (\eq{BC})
are not defined for $\kp<0$. At first glance it seems to be
awkward and the question immediately arises whether a sensible
continuum limit can be obtained in this region. The fact that for
small positive $\kp$ and for negative $\kp$ there is
a broken symmetry phase FM(W)
at large enough Yukawa coupling suggests that the Yukawa coupling
must be the driving force for SSB in that region
(see also ref.~\cc{Cah91}).
This is obviously outside the regime of usual
perturbation theory in the continuum. It is therefore very important
to investigate this region if one is looking for nonperturbative
effects of the Yukawa coupling.

Of course, the AM and FI phases appearing only for $\kp < 0$
are probably lattice artefacts as they depend very much
on the lattice geometry.
But the scaling regions of the FM(W) and PMW phases down to the
point A (fig.~1) are worth of consideration.

The Osterwalder-Schrader (OS) reflection positivity
can be proven presumably only for $\kp \ge 0$.
However, it is a sufficient, but not a necessary
condition for unitarity, so that unitarity can still hold.
In the numerical computations of the propagators of the theory, our
failure to detect any state with a negative norm is assuring.
Furthermore, as demonstrated later in this article and also
in ref.~\cc{BoDe91a},
the measured values for all the masses and the renormalized
couplings continue analytically from positive to negative $\kp$
across $\kp=0$.

The problems with the transition from $\kp \ge 0$ to $\kp < 0$
seem to exist only on the level of bare parameters in the
continuum parametrization. One should consider only the
renormalized quantities. If one knew the renormalized running
coupling
$\overline{y}(\mu)$ at all momentum scales $\mu$
one could define also a sensible bare Yukawa coupling
$y_B$ as
\be
y_B=\yb(\mu \sim 1/a)\;.  \lb{YYYY}
\ee
For small
values of $y_0$ and $g_0$, usual perturbation theory is valid and
$y_B$ would not differ very much from the
coupling parameter $y_0$. At larger values of $y_0$ and in particular
in the negative $\kappa$ region the parameter $y_0$ does not any longer
have to be close to $y_B$. This so-defined bare coupling $y_B$ would
also have
in general no simple relation to the bare parameters $\kappa$ and $y$.
Fig.~3 schematically illustrates how $y_B$ and $y_0$
may split up from each other as the nonperturbative region is entered
along some line of constant physics specified, e.g.,
by $y_R = \overline{y}(\mu=\mu_{\rm phy}) = const$, where
$\mu_{\rm phy}$ is a physical scale.

The stage is therefore well set to plunge into the negative $\kp$ region
with three issues in mind:
\begin{itemize}
\item[(i)] Is there a nonperturbative fixed point?

\item[(ii)] Even if the theory is
trivial, can the couplings be strong at a reasonably large cut-off?

\end{itemize}
Our previous investigation \cc{BoDe91a} with more naive methods of
analysis of the numerical data has already produced
tentatively negative answers to
these two questions which we want to confirm in this paper.
\begin{itemize}
\item[(iii)] In any case,
it is necessary to find out how far the observables
in the $\kp < 0$ region differ {\em quantitatively} from
the $\kp > 0$ region, for determination of bounds on
renormalized couplings.
\end{itemize}

In the rest of this section we want to point out the connection
of a Yukawa theory with a Nambu-Jona-Lasinio (NJL) type model.
Integrating out the scalar fields in the partition function
defined by the action (\eq{SS}) with $\kp = \lambda = 0$
produces obviously a pure fermionic theory with local (on-site)
four-fermion coupling of strength $\half y^2$ -- the
NJL model on the lattice.
The scalar field is equivalent to the ``auxiliary field'' used
in the context of the four-fermion coupling
\cc{Na88,BaHi90,GrNe74}.
Also for $\lambda > 0$ the effective fermion interaction
is local at $\kp = 0$ and corresponds thus to a multifermion interaction
discussed, e.g., in ref.~\cc{Su90}.
Thus we conclude that theories of the NJL type are special cases
($\kp = 0$) of the Yukawa models in the lattice parametrization.
But in terms of the renormalized theory, as we have discussed
above, $\kp=0$ is not singular and the same qualitative physics is
obtained for the whole FM(W)-PMW scaling region down to the point A
unless there is a nonperturbative fixed point somewhere. Recent
work \cc{HaHa91,Zi91} using $1/N$ expansion also shows an equivalence
between Yukawa models and generalized NJL models.

Thus we achieve a unification of concepts and language
of the SSB: the Yukawa models treated nonperturbatively using the
lattice formulation (\eq{SS})
embrace both
\begin{itemize}
\item the classical Higgs mechanism, in which the SSB
      is understood in terms of the quasi-classical approximation
      for the effective potential of the scalar field
      and perturbation theory
      (e.g. the region $y \ll 1$ in fig.~1 for any $\lambda$), and
\item the NJL type mechanism, operating at small
      (and possibly negative) $\kp$
      for relatively large values of $y$,
      which has to be treated by nonperturbative techniques
      such as the $1/N$ expansion.
\end{itemize}
Yukawa theories provide a gradual transition between these mechanisms.
It seems, therefore, possible to formulate the SSB in the standard model
in terms of the NJL mechanism \cc{Na88,BaHi90,Su90}. However,
in the light of the above discussion, the distinctions between
an elementary and a composite scalar, an auxiliary and a dynamical
scalar field and between dynamical symmetry breaking  and the usual Higgs
mechanism do not seem important.

\section{Properties of the scalar propagators}

In our numerical simulation
we consider $V=L^3T$ lattices with periodic
boundary conditions for the scalar fields. Fermionic fields are
periodic in space and antiperiodic in euclidean time directions.

In the {\bf symmetric (PM) phase} the bosonic spectrum contains
the $\Phi$ quadruplet of mass $am_{\Phi}$.
The corresponding propagator in the momentum space is
\be
G_{\Phi}(a p) = \left\lag \frac{1}{4V} \sum_{x,y} \htr
\{ \Phi_x^{\dagger} \Phi_y  \}
\exp( i a p (x-y) ) \right\rag \;. \lb{bp}
\ee
Neglecting the instability of $\Phi$ at the present
precision level,
the renormalized mass $am_{\Phi}$
and the wave function renormalization
constant $Z_{\Phi}$ can be defined by means of the limit
\be
G_{\Phi} (ap)|_{p^2 \ra 0} = \frac{ Z_{\Phi} }{(am_\Phi)^2 + \hw^2} \;,
\lb{bpp}
\ee
where the quantity $\hw^2=2 \sum_{\mu} (1-\cos(ap_{\mu}))$ is the
dimensionless lattice equivalent of the momentum square in the
continuum.

In the {\bf broken (FM) phase} it is useful
to introduce the following notation for the scalar field
\be
\Phi = \sigma \un + i \sum_{j=1}^{3} \pi^j \; \tau^j \;.
                                                        \lb{SIGPI}
\ee
Here $\tau^j$, $j=1,\ldots,3$ are the usual Pauli matrices and
$\sqrt{ \sigma_x^2 + \sum_{j=1}^3 (\pi_x^j)^2 }=1$. The
components are chosen such that the magnetization is given by
$\lag \Phi \rag  = \lag \sigma \rag$.
Then the longitudinal component    is
associated with the massive $\sigma$ boson whereas the transverse
ones with the three massless Goldstone bosons $\pi$.
On the lattice the $\pi$
and $\sigma$ propagators in the momentum space are defined by
\bea
G_{\pi}(a p) &=& \left\lag \frac{1}{3V} \sum_{x,y}
\sum_{j=1}^{3} \pi_x^j \pi_y^j
\exp( i a p (x-y) ) \right\rag \;,  \lb{gp} \\
G_{\sigma}(a p) &=& \left\lag \frac{1}{V} \sum_{x,y}
\sigma_x \sigma_y
\exp( i a p (x-y) ) \right\rag \;. \lb{sp}
\eea
The only asymptotic states in the FM phase are the massless $\pi$
bosons.
Therefore the wave function renormalization constant $Z_{\pi}$
for the scalar field is defined through the following limit of the
$\pi$ propagator,
\be
G_{\pi}(ap) |_{p^2 \ra 0} = \frac{Z_{\pi}}{\hw^2}\;. \lb{gpp}
\ee
Using the so-defined $Z_\pi$ the renormalized field expectation value
$v_R$ is then given by eq.~(\eq{vR}).
Again at our present precision level it is presumably sufficient
to define the renormalized mass $am_{\sigma}$ of the unstable
$\sigma$ particle by the relation
\be
G_{\sigma} (ap)|_{p^2 \ra 0} = \frac{ Z_{\sigma} }
                                    { (am_{\sigma})^2 +\hw^2 }
\;. \lb{spp}
\ee
Another point to note is
that in the pure $\Phi^4$ theory the renormalized mass
defined this way is very close to the physical mass
\cc{KuLi88b,LuWe}.

In a finite system no spontaneous breakdown
of the symmetry can occur and
during a Monte Carlo simulation
in the broken phase
the system drifts through the set of degenerate ground states.
This causes a vanishing of
noninvariant observables like $\vv$.
To compensate for this drift,
each scalar field configuration is rotated so that
$\frac{1}{V} \sum_x \Phi_x = \frac{1}{V} \sum_x \sg_x$.
In the pure $\Phi^4$ theory the rotation technique provides a very good
approximation of the infinite volume values of the noninvariant
quantities \cc{Ha}.

A lot is known about the properties of the scalar propagators in the pure
$\Phi^4$ theory which is the limiting case
of the model (\eq{SS}) both for $y=0$
and $y=+\infty$
(in the latter case fermions become infinitely heavy and
decouple completely from the particle spectrum).
When plotting the inverse propagators
$G_{\Phi}^{-1}(ap)$, $G_{\pi}^{-1}(ap)$ and $G_{\sigma}^{-1}(ap)$
as functions of $\hw^2$ one finds for all propagators
and for all possible values of $\hw^2$ a straight line behaviour confirming
the analysis of the data in terms of free scalar propagators.
{}From straight line fits to the inverse propagator data,
using the relations (\eq{bpp}),
(\eq{gpp}) and (\eq{spp}), one can determine
the wave function renormalization constants
and the renormalized masses on a finite lattice.

In the Yukawa model we have determined
for various values of $\kp$ and $y$ the momentum space
propagator (\eq{bp}) in the PMW and PMS phases and the propagators (\eq{gp})
and (\eq{sp}) in the FM phase close to the critical lines FM(W)-PMW and
FM(S)-PMS (see fig.~1).
For very small and very large values of the Yukawa coupling
$y$ the results for the propagators are very close to
those obtained in the $\Phi^4$ theory, i.~e., when plotting the inverse
propagators as functions of $\hw^2$ we find approximately a straight
line behaviour.
However, when entering the intermediate Yukawa coupling region
and lowering $\kp$,
where fermions have a strong feedback on the scalar sector
and the staggered scalar states (sect.~3) become visible,
deviations from the free propagator behaviour at larger values
of $\hw^2$ are observed.
These deviations become more and more pronounced
when approaching the multicritical points A and B.

In fig.~4 we display for several points in the FM phase the inverse
Goldstone propagator $G_{\pi}^{-1}(ap)$
as a function of the quantity $\hw^2$. The
figures in the left column were obtained at 3 points in the vicinity
of the FM(W)--PMW phase transition whereas the figures in the right
column correspond to 3 points in the vicinity of the FM(S)--PMS
phase transition.
The lowest figures were obtained very close to the
multicritical points A and B respectively.
The figures show that there are three kinds of
deviations from a free propagator behaviour:
\begin{enumerate}
\item  The formation of a second pole in $G_{\pi}(ap)$
       in the corner of the Brillouin zone with $ap=(\pi,\pi,\pi,\pi)$
       when
       approaching the points A or B within the FM phase. This effect
       is caused by the staggered states $\Phi^{st}$
       which are present on the lattice
       also in the FM phase (see sect.~3).

\item  The appearance of dips in $G_{\pi}^{-1}$ around the momenta
       $\hw^2=4,8,12$ (corresponding to $a p_\mu \!=\! 0$ or $\pi$)
       in the weak coupling region. These dips
       are already visible at positive $\kp$ as can be seen from the
       first figure in the left column.

\item  At small $\hw$ the inverse propagator has in the weak Yukawa
       coupling region a curvature, which plays a significant
       role in the data analysis.
\end{enumerate}
Similar structures were also discovered
for the propagator $G_{\sigma}(ap)$
in the FM phase and the propagator $G_{\Phi}(ap)$ in the
PMW and PMS phases near the FM(W)--PMW and FM(S)--PMS phase transitions.
As we shall now discuss,
the first two effects are actually lattice artefacts,
dependent on the geometry of the lattice,
but they have to be understood quantitatively
in order to extract the physically relevant quantities
from the scalar propagator data.
The third effect is physical.


\section{Staggered scalar states in the FM and PM phases}

For an understanding of the two pole phenomenon it is useful to
discuss first the situation in the pure $\Phi^4$ theory
which is found in the limiting cases $y=0$ and $y=+\infty$.
We use the $\sigma$ propagator
$G_{\sigma}(ap,\kp)$ as an example and indicate for a moment
explicitly the $\kappa$-dependence of the propagator.
Using the transformation (\eq{ST}) we find the relation:
\bea
G_{\sigma}(ap,-\kp) =
       G_{\sigma}(ap+a\tilde{\pi},\kp)
\lb{spst}
\eea
where $\tilde{\pi}=(\pi,\pi,\pi,\pi)$.
As the propagator $G_{\sigma}(ap,\kp)$ in the pure scalar theory
can for all momenta $ap$
be well described by the free scalar Ansatz eq.~(\eq{spp}),
$G_{\sigma}(ap,-\kp)$ has for $\kp > \kp_c$ the form
\be
G_{\sigma}(ap,-\kp) =
 \frac{ Z_{\sigma} }
 { (am_{\sigma})^2 +\widehat{ a(p+\tilde{\pi}) }^2 }
= \frac{ Z_{\sigma}^{st} }
 { (am_{\sigma}^{st})^2 + 16 - \widehat{ ap }^2 }
 \lb{sppst}
\ee
with $m_{\sigma}^{st}=m_{\sigma}$ and $Z_{\sigma}^{st}=Z_{\sigma}$.
For $am_{\sigma}=0$
the propagator $G_{\sigma}(ap,-\kp)$ has thus a pole at the momentum
$p=\tilde{\pi}$.
Similar relations can be obtained also for the propagators
$G_{\Phi}$ and $G_{\pi}$.
Obviously, at no $\kp$ the normal and staggered states appear
simultaneously as their masses are small only close to the
FM--PM and PM--AM phase transitions, respectively, which are distant.

For $0 < y < \infty $ the symmetry (\eq{ST}) does not hold any more.
Nevertheless, for small and large values of $y$ the spectrum is
very similar to the pure $\Phi^4$ theory.
But when the phase transition lines
approach each other and finally meet at the points A and B
we expect that both the normal and the staggered state masses are
small simultaneously.
In particular, $am^{st}_{\Phi}$ can be small also in
the FM phase.
In the FM(S) phase it is therefore reasonable to try to
fit the numerical results for $G_{\sigma}(ap)$
by the two pole Ansatz
\be
G_{\sigma}(ap) =
\frac{Z_{\sigma} }{(am_{\sigma})^2+\hw^2 } +
\frac{Z_{\Phi}^{st} }{(am_{\Phi}^{st})^2+16-\hw^2 } \;.\lb{stan}
\ee
Analogous expressions have been applied also for the propagators
$G_{\pi}(ap)$ and $G_{\Phi}(ap)$ in the
FM and PM phases respectively.
In fig.~5 we show as an example a fit to the Goldstone
propagator in the FM(S) phase,
where the fit Ansatz for $G_{\pi}(ap)$
is given by eq.~(\eq{stan}) with
$am_{\pi}=0$ and $Z_{\sigma}$ replaced by $Z_{\pi}$.
Fig.~5 shows that the
data are described by the two pole Ansatz very well.
We furthermore expect
that the staggered mass $am_{\Phi}^{st}$ and the
wave function renormalization constant $Z_{\Phi}^{st}$ obtained
from the fits
to $G_{\pi}(ak)$ and $G_{\sigma}(ap)$ should agree.
This expectation is indeed confirmed by the numerical
results, for example at $\kp=-0.65$ and $y=1.8$ the
values are $am_\Phi^{st}=1.47(7)$ / $1.37(4)$ and
$Z_\Phi^{st}=1.41(7)$ / $1.45(4)$
from the $\sigma / \pi$ propagators.
As expected, the mass $am_{\Phi}^{st}$ does approach zero when the
line AB is approached. It should be also stressed
that both terms in the Ansatz (\eq{stan}) have positive sign, so that the
second term has the usual form of a pole after shifting the
momentum by
$(\pi,\, \pi,\, \pi,\, \pi)$
and cannot be interpreted as a ghost.

The same formulae describe also
the two pole structure of the scalar propagator in the
vicinity of the point A. Here, however,
a more elaborated Ansatz has to be
developed in order to take into account simultaneously also
the overlaid finer structure caused by the
fermion loop corrections.
It is the subject of the next section.

\section{One fermion loop contribution to the scalar propagators}

The other two features
making the scalar propagators different from a free propagator are
the appearance of dips at momenta $\hw^2 \!=\! 4$, $8$, $12$
and the curvature at
small $\hw^2$.
They occur in the weak coupling phase regions FM(W) and PMW
where the fermion masses scale.

According to the definitions of the quantities
$Z_{\pi}$, $Z_{\sigma}$ and $am_{\sigma}$ in eqs.~(\ref{gpp}) and
(\ref{spp}) the scalar propagators have to be analyzed
in the limit $\hw^2 \ra 0$.
However, on small lattices with periodic boundary conditions
the smallest nonvanishing momentum is
$ap=\frac{2\pi}{T}$ (for $T>L$)
which is quite large, e.g., $ap=0.79$ for $T=8$.
In the pure $\Phi^4$ theory this is not a serious problem
as the inverse scalar propagator can be fitted with a
straight line up to $\hw^2 = O(10)$ \cc{KuLi88b}.

In our Yukawa model the situation is much less favourable:
in addition to the
dips occuring at $\hw^2=4, \; 8$ and $12$ -- for which
one might argue that they should
simply be ignored as the analysis has to be
restricted to the smallest momenta --
we have to face the more serious problem of a significant curvature
at small $\hw^2$.
The following further observation makes the situation look even worse:
Increasing the $T$-extent of the lattice -- which is the cheapest
possibility to have small momenta --  e.g. from
$T=6$ up to $T=46$, we still have not found an onset of a linear
$\hw^2$ dependence.
We conclude from this that
for sizeable Yukawa coupling
an application of
the free particle Ansatz for the scalar propagators
on finite lattices
produces uncontrollable systematic errors.

Therefore, we have developed a more sophisticated
fit Ansatz for the scalar propagator
based on the 1-fermion-loop contribution to the
self-energy of the $\pi$ or $\sigma$ bosons.
The justification for including only the fermion loop
comes from the experience in the pure
scalar sector of the model where the scalar
loop contributions do not change the linear $\hw^2$-shape
of the propagators but only lead to wave function and mass
renormalizations.
On the other hand, as will be shown below,
in the case of naive lattice
fermions the 1-fermion-loop contribution
will cause deviations from the linear $\hw^2$-dependence
just of the form observed in the data.

Let us discuss the example of the Goldstone boson propagator.
On finite lattices we may write
\be
G_{\pi,L}^{-1}(ap)=Z_\pi^{-1} \left[(am_{\pi,L})^2+
\widehat{ap}^2
     - \Sigma_{\pi,L} (ap;am_{F,L}) \right] \; ,   \lb{GL}
\ee
where $\Sigma_{\pi,L}$ denotes the 1-fermion-loop contribution to the
self-energy of the Goldstone boson
in the renormalized perturbation theory.
For the purpose of this section the subscript $L$ points out
 the possible dependence of various
quantities on the spatial lattice size $L$.
For instance with regard to the Goldstone bosons
we take the finite-size of the lattice into account
in a naive but simple way -- allowing for a finite mass of the
Goldstone boson $am_{\pi,L}$ (see e.g. also \cc{AoShi91}).
(A treatment based on chiral perturbation theory,
successful in the pure $\Phi^4$ theory \cc{Ha}, is in the
complex situation with light fermions
presumably not applicable.)
In the FM phase where the fermions are massive
we impose the two necessary normalization conditions
on $\Sigma_{\pi}$ in the infinite volume at momentum zero,
\bea
\left. \Sigma_{\pi,\infty} (ap;am_{F,\infty})
\right|_{p=0} &=& 0 \; ,  \nonumber \\
\left. \frac{\partial}{\partial \widehat{ap}^2} \, \Sigma_{\pi,\infty}
(ap;am_{F,\infty}) \right|_{p=0} &=& 0 \; .     \lb{normcond}
\eea
With this normalization $G_{\pi}$ approaches in the thermodynamic limit
the form used for the definition of $Z_{\pi}$, eq.~(\ref{gpp}), provided
$(am_{\pi,L})^2 \rightarrow 0$ as $L \rightarrow \infty$.

The not-yet-normalized
$\Sigma_{\pi,L}^\prime$ calculated from
the corresponding Feynman diagram on the lattice is
\be
\Sigma_{\pi,L}^\prime (ap;am_{F,L}) =
(-1) \frac{4}{L^3 T} \sum_k \mbox{Tr } \left\{
  (i \gamma_5 \tilde{y}_R) \frac{1}{i \ssl(k)+am_{F,L}}
  (i \gamma_5 \tilde{y}_R)
  \frac{1}{i \ssl(k-ap)+am_{F,L}} \right\} \; .   \lb{loop}
\ee
This equation defines the renormalized Yukawa coupling
$\tilde{y}_R$ in terms of the 3-point function.
Of course, $\tilde{y}_R$ could in principle differ from
$y_R$ defined in eq.~(\ref{TR}).
In the above $s_{\mu}(k)=\sin k_{\mu}$,
the factor $(-1)$ comes from the fermion loop,
the trace is to be taken over the Dirac indices.
The SU(2) trace
together with another factor of 2 from the HMC doubling
(see sect.~2) results in the factor 4
whereas the standard fermion doubling is taken into account
automatically.
The sum over the loop momentum $k$
runs over the set of momenta corresponding to a finite $L^3 T$ lattice
with periodic boundary conditions in the space direction and
antiperiodic boundary conditions in the time direction,
\be
\begin{array}{rlll}
k_j &= \frac{2 \pi}{L} n_j \; , &\hspace{0.2cm} n_j=0,1,2, \ldots L-1
 &\hspace{0.3cm} j=1,2,3  \;\; , \nonumber  \\
k_4 &= \frac{2 \pi}{T}(n_4+\frac{1}{2}) \; ,
                                 &\hspace{0.2cm} n_4=0,1,2, \ldots T-1
   \; . &     \lb{moms}
\end{array}
\ee

After some simplifications we get
\bea
\Sigma_{\pi,L}^\prime (ap;am_{F,L}) &=&
 \tilde{y}_R^2 \frac{16}{L^3 T} \sum_k
\frac{(am_{F,L})^2 + s(k)s(k-ap)}
{ \left[ (am_{F,L})^2 + s^2(k) \right]
\left[ (am_{F,L})^2 + s^2(k-ap) \right]}  \nonumber \\
 &=:& \tilde{y}_R^2 \, I_{\pi,L}^\prime (ap;am_{F,L})\; ,
\lb{simpl}
\eea
where we introduce the notation $I_{\pi,L}^\prime$
for the (unnormalized)
lattice integral after factorizing out $\tilde{y}_R^2$.
In order to recognize the 1-fermion-loop as the reason
for the deviations in the scalar propagators
it is instructive to discuss the dependence of this expression
on the external momentum $ap$.
For $ap=(0,0,0,0)$ the denominator has
minima when the loop momentum
components are $k_{\mu}\!=\!0$ or $\pi$, corresponding
to processes involving the physical fermion and its antifermion
or some doubler fermion with its own antidoubler.
Thus a considerable contribution of $\Sigma_\pi$ to $G_{\pi}$ for
small momenta $ap$ can be expected, causing the observed
strong curvature of $G_\pi^{-1}$.
In addition, $\Sigma_\pi$ also peaks
when
some components
$ap_{\mu}\!=\!\pi$ and others are zero.
Then the kinematics allows
such intermediate states to be excited
which involve e.g. the physical fermion and the
antidoubler of momentum $ap$,
or any other appropriate pair of doubler and antidoubler
whose respective positions of poles differ just by $ap$.
This is precisely the reason for the dips seen in the inverse
scalar propagators near $\hw^2=4, \, 8, \, 12$.

To fit the propagator data using the Ansatz (\ref{GL})
we have to perform the following steps:
First, $\Sigma_{\pi,L}$ is normalized as
\bea
  \Sigma_{\pi,L}(ap; am_{F,L})
  &=& \Sigma_{\pi,L}^\prime (ap;am_{F,L})
    - \Sigma_{\pi,\infty}^\prime (ap=0;am_{F,\infty})
                         \nonumber \\
  & &-\hw^2 \left( \left. \frac{\partial}{\partial \hw^2}
                  \Sigma_{\pi,\infty}^\prime
             (ap; am_{F,\infty}) \right|_{p=0} \right)
\lb{normed}
\eea
so that it satisfies the conditions (\ref{normcond})
in the infinite volume limit.
Here the fermion mass $m_{F,L}$ on the given finite lattice
is taken from the standard fit to the fermion propagator data.
But note that this normalization also requires an
estimate of the fermion mass in infinite volume $am_{F,\infty}$
(various attempts to normalize $\Sigma_{\pi,L}$ using only
finite volume quantities did not work).
In the spontaneously broken phase FM, the major part of
the finite-size effects is expected to be due to the massless
Goldstone bosons leading to a volume dependence linear
in $1/L^2$.
So, checking this dependence and then extrapolating
$a m_{F,L}$ to $a m_{F,\infty}$
requires, at a given $(\kappa, \, y)$-point,
simulations on a sequence of at least three lattices.
We have performed runs on lattices $L^3 16$ with
$L=6, \, 8, \, 10$ and $12$ with the result that
as long as
$a m_{F,L}$ itself is not too small the agreement with a linear
$1/L^2$ dependence indeed
allows an extrapolation to $a m_{F,\infty}$
(see next section).

Analogous to eq.~(\eq{simpl}) we define the (normalized)
lattice integral $I_{\pi,L}$ with $\tilde{y}_R^2$ factorized out,
\be
  \Sigma_{\pi,L}(ap; am_{F,L}) =:
  \tilde{y}_R^2 I_{\pi,L} (ap; am_{F,L})   \; .
\lb{defI}
\ee
The $\pi$ propagator fit Ansatz is then
\be
G_{\pi,L}^{-1}(ap) =
                     Z_\pi^{-1} \left[ a m_{\pi,L}^2+\hw^2
     - \tilde{y}^2_R \, I_{\pi,L}(ap;a m_{F,L}) \right]
       \lb{ANSATZ}
\ee
with the free parameters
$a m_{\pi,L}$, $Z_\pi$ and $\tilde{y}_R$
(we note that through the normalization conditions (\eq{normcond})
this expression also depends on $m_{F,\infty}$).

Before describing the results
in the next sections
let us discuss the quality of the fit.
The full Ansatz we use is the superposition of
the propagator with the pole at $ap=(0,0,0,0)$
including the 1-fermion-loop
contribution and the staggered propagator with the pole at
$ap=(\pi,\pi,\pi,\pi)$ as explained in sect.~6 above.
This Ansatz is able to describe the scalar propagator data
in the complete interval $0 \leq \widehat{ap}^2 \leq 16$.
In particular, the curvature at small momenta and
also every detail of the peculiar structures near
$\widehat{ap}^2=4, \, 8, \, 12$
are perfectly reproduced. This is demonstrated
in figs.~6 and 7 for two
typical examples of the fits, one at small positive $\kappa$
and one in the close vicinity of point A
where the scalar propagators have the most complex form.

It should be stressed again that the dips at momenta
$\widehat{ap}^2=4, \, 8, \, 12$ are caused by the presence of
the doubler fermions and hence should at least look different,
if not
absent, in models without them.
However, in all models the curvature at small momenta
will appear for sufficiently strong Yukawa coupling
and the second pole at
$\widehat{ap}^2=16$ will be present near
phases with antiferromagnetic ordering.

\section{Fermion and scalar masses at small positive $\kp$
         and negative $\kp$}

We have been able to determine reliably both $am_F$ and $am_\sg$
simultaneously only for $\kp \geq 0$.
Here $am_\sg$ is always greater than $am_F$  at least by a factor 3 -- 6.
This presumably is a consequence of having a large
number (32) of fermion doublets.
Estimating that a similar mass-ratio holds also for $\kp < 0$
we have mostly performed calculations at points with very small
fermion mass, $am_F \simeq 0.1 - 0.3$,
in order to have the $\sg$-boson mass at least smaller than 1.

The determination of $am_F$ in this range of values requires only
moderate statistics and can be reliably performed
at $\kp < 0$ even in the vicinity of the point A.
An important condition is, however, that the long size $T$ of the
lattice $L^3T$ is at least 16, otherwise $am_F$
is spuriously small and $T$-dependent when determined from the fit
by means of the free fermion propagator.
The method of analysis and many results have been presented
already in ref.~\cc{BoDe91a}.
Here we would like to point out the large spatial volume dependence
of $am_F$.
For $am_F \simeq 0.3$ the value can decrease by 40\%
when $L$ increases from 6 to 10.
Nevertheless, for $am_F \gsim 0.2$ the decrease is linear with $L^{-2}$,
so that one can tentatively extrapolate to $L = \infty$.
The observable $\vv$ is easily measurable everywhere and
also has a linear $L^{-2}$ dependence.
The $L^{-2}$ dependence and the extrapolation of both the quantities
are shown in fig.~8 (we show $av_R$ instead of $\vv$ because $Z_{\pi}$
is $L$-independent as discussed in sect.~9).

The $\sg$ boson mass has been determined at several points
for $\kp \geq 0$ on lattices of various sizes.
The finite-size effects are compatible
with the expected $L^{-2}$ dependence
but the large error bars prevent a verification.
Nevertheless, we have used the same volume dependence to extrapolate
$am_\sg$ to $L =\infty$.

There are two technical reasons
making a reliable determination of $am_\sg$ at negative $\kp$ very
difficult.
Firstly, the number of iterations for
the fermion matrix inversion needed for the field update
increases drastically
with decreasing $\kp$.
As the determination of the $\sg$ propagator requires much higher
statistics than of the fermionic propagator,
an accumulation of good data
in the negative $\kp$ region, in particular
in the vicinity of the point A is prohibitively expensive.
Secondly, the maximum of the curve $G_{\sigma}^{-1}(ap)$ occurs already
at rather small momenta (see fig.~7) and
only a few data points of the propagator contain
the information about $am_\sg$ and $Z_\sg$,
the rest being dominated by the staggered state.
Therefore we have not succeeded to determine reliably $am_\sg$
for any of our data points at negative $\kp$.


The largest renormalized couplings are expected on the boundary of the
scaling region, i.e. relatively far from the critical line.
However, at present we do not know the position of this boundary,
actually not even its proper definition (e.g. how small
$am_F$ and $am_\sg$ should be in order that
the lattice model can be used as an effective continuum theory).
We are thus not able to extract upper bounds on masses from our
results for renormalized couplings.

\section{The renormalized Yukawa coupling}

The excellent agreement of the fits with the MC data for the
$\pi$ propagator both for positive and negative $\kp$
allows to determine $Z_\pi$ reliably.
In comparison to the usual determination of $Z_\pi$
by a naive free particle fit to the smallest
momentum used e.g. in our earlier publication \cc{BoDe91a}
the present method yields more precise results
(the error bars are reduced by factors 3 -- 10).
In particular, they are now stable
when the $T$-size of the lattice is varied, in spite
of the strong curvature near $\hat{p}^2=0$ which is most
clearly seen on large-$T$ lattices.
No $L$-dependence of $Z_\pi$ has been found.
The actual values are slightly smaller
than found in ref.~\cc{BoDe91a}
confirming the conjecture in that paper
that a simple linear fit to the smallest
momentum on small lattices gives overestimated values of $Z_{\pi}$.

Some values of $Z_{\pi}$
obtained in the vicinity of the FM(W)--PMW critical line
are plotted in fig.~9.
$Z_\pi$ decreases strongly as $\kp$ decreases and
appears to vanish at the multicritical point A.

The knowledge of $Z_{\pi}$ allows us to determine, from the
available very good data for $am_F$ and $\vv$ for various $L$,
the renormalized Yukawa coupling $y_R$ by means of
the definition (\eq{TR}).
Both $am_F$ and $\vv$ are strongly $L$-dependent,
but their ratio turns out to be practically $L$-independent.
We use the observed linear $L^{-2}$ dependence to
extrapolate $am_F$ and $av_R$ to infinite volume
obtaining $y_R$ in the $L \rightarrow \infty$ limit (see fig.~8)
In fig.~10 we plot these results against
the fermionic correlation
length $\xi_F=1/am_F$ at $L \!=\! \infty$.
The reason for not using the smaller $\xi_{\sigma}=1/am_\sg$
at $L \!=\! \infty$
for this purpose is the fact that we do not know its values for $\kp < 0$.

The dotted curve in fig.~10, given by
\be
y_R=\frac{1}{\sqrt{\frac{32}{4\pi^2} | \ln a \mu | }} \; ,
\lb{dotc}
\ee
is obtained by choosing infinite bare Yukawa coupling in the
1-loop formula for the running Yukawa coupling
and identifying the scale $\mu$ to be $m_F$.
As pointed out in \cc{BoDe91a},
this curve described quite well the results for $y_R$
at small positive $\kp$ and negative $\kp$.
The dotted error bars associated with this
curve in the figure indicate the range of values and the error bars
of those former results in \cc{BoDe91a}.

The dramatic reduction of the error bars in fig.~10
is mostly due to the refined analysis in this paper.
It is now apparent that $y_R$ is bounded by the dotted curve,
suggesting applicability of 1-loop perturbation theory
even close to the point A. In addition all the $y_R$ results
are clearly below the $s$-wave tree level unitarity bound
for 32 fermion doublets.
This bound is indicated by the horizontal dashed line
in fig.~10.
These results suggest the conclusion that in our model there is
no strong Yukawa coupling regime at least for $\xi_F > 5$.
We remark that initial indications from numerical investigations
in the mirror fermion model \cc{MIRRORS} are, however, different.
The weakness of $y_R$ in our model in the negative $\kp$ region
comes about as follows:
the unrenormalized ratio $a m_F / \vv$ actually increases
strongly as $\kp$ decreases, but this is compensated by
the equally strong decrease of $Z_\pi^{1/2}$ (fig.~9).

The fermion loop correction allows to determine
the renormalized Yukawa coupling $\tilde{y}_R$
defined in terms of the
vertex function and extracted from the Goldstone propagator data
by the Ansatz (\ref{ANSATZ}).
In table~\ref{tab:yReta} we compare $y_R$ and $\tilde{y}_R$
for some typical $(\kappa,y)$-points where we were able
to extrapolate $am_F$ to $L=\infty$ in order to
satisfy the normalization conditions (\ref{normcond}).
\begin{table}
\begin{center}
\begin{tabular}{|c|r|l|l|l|cl|r|} \hline \hline
             &Lattice &\k=0.03 &\k=0.04 &\k=-0.65& \phantom{.}
             &Lattice&\k=0.00  \\
             &        & y=0.60 & y=0.60 & y=0.98 &
             &       & y=0.65  \\
\hline \hline
$    y _R   $&$6^3 24$&        &        &0.427(5)& & $6^3 24$& 0.55(2) \\
$\tilde{y}_R$&        &        &        &0.49(4) & &         & 0.51(1) \\
\hline
$    y _R   $&$6^3 16$&0.410(6)&0.461(8)&0.412(6)& & $6^3 12$& 0.55(2) \\
$\tilde{y}_R$&        &0.401(4)&0.45(1) &0.495(8)& &         & 0.51(2) \\
\hline
$    y _R   $&$8^3 16$&0.410(6)&0.47(2) &0.43(1) & & $8^3 12$& 0.51(1) \\
$\tilde{y}_R$&        &0.402(6)&0.450(4)&0.49(2) & &         & 0.52(1) \\
\hline
$    y _R   $&$10^316$&0.40(1) &0.46(1) &0.45(3) & &$10^3 12$& 0.51(3) \\
$\tilde{y}_R$&        &0.402(5)&0.45(1) &0.48(7) & &         & 0.52(4) \\
\hline \hline
\end{tabular}
\end{center}
\caption{{\em Comparison of $y_R$ from the tree level definition
({\protect\ref{TR}}) to $\tilde{y}_R$ from the fit to the
Goldstone propagator ({\protect\ref{ANSATZ}})
at some typical $(\kappa,y)$-points.}}
\label{tab:yReta}
\end{table}
The good agreement
found between $y_R$ and $\tilde{y}_R$
indicates that the analysis of the scalar propagator data
by taking the 1-fermion loop corrections into account is adequate
and that both definitions of the renormalized
Yukawa coupling quantitatively agree.
This further supports the conclusion that in our model with naive
fermions the Yukawa coupling is not strong.

Some caution is due, however.
It could be that our linear extrapolation of $a m_F$ to
$L \!=\! \infty$ underestimates $\xi_F$. Furthermore, as
we do not know $am_\sg$ in the vicinity of the point A,
we cannot exclude that $\xi_\sg$ is as large as $\xi_F$ and that
our data points there are actually deep in the scaling region.

\section{Fermion influence on the scalar mass bound}

To study the influence of heavy fermions on the triviality bound
for the scalar mass it would be again most interesting
to perform this analysis in the vicinity of the multicritical
point A and the lack of reliable results
for $am_\sg$ there is deplorable.
However, provided the renormalized Yukawa coupling does not
attain in the negative $\kappa$ region
values significantly larger than at $\kp$ positive,
as suggested by our results,
we expect that it is sufficient to investigate
the influence of the Yukawa coupling
on the scalar sector in the positive $\kappa$ region.

These considerations have motivated  our relatively high statistics
study of the $\sg$-pro\-pa\-ga\-tor at $\kappa \gsim 0$.
We have fixed $y$ at the value $y=0.6$, for which the critical $\kappa$
is $\kappa_c=0.020(5)$.
On lattices of three different spatial sizes, $L^3 16$ with $L=6,8,10$,
we have accumulated about 4-5 thousand Hybrid Monte Carlo trajectories
at three points $\kappa=0.03,0.04,0.06$.
Both scalar propagators have been analyzed by means of the
same Ansatz (\eq{ANSATZ}) and the scalar mass $am_{\sigma}$ and
$Z_{\pi}$ have been determined.

The results for the ratio $m_\sg/v_R$
are displayed in fig.~11 as a function of the scalar correlation length
$\xi_{\sigma}=1/am_{\sigma}$ (open symbols).
The finite-size dependence of $a m_\sigma$ can be described to be
linear in $L^{-2}$, though the error bars on $a m_\sigma$
leave room for modification.
The tentative extrapolation of the results to the infinite volume
assuming a $L^{-2}$ dependence
is indicated by the full squares.
This kind of plot is customary to extract
a triviality upper bound for the
scalar mass in the pure $\Phi^4$ theory.
For comparison the data from this
theory ($y=0$) on a hypercubic lattice \cc{O4uppb} are also shown
(full circles).

The results in the Yukawa model
on finite lattices approach the infinite volume results
from below
(different from the $\Phi^4$ theory \cc{Ha})
and the ex\-tra\-po\-lated results are - within large error bars -
consistent with the pure $\Phi^4$ results.
As $\xi_\sigma \simeq 1$ -- $3$ and $\xi_F$
is much larger than $\xi_\sigma$
for all points in fig.~11, one can expect
that the edge of the scaling region is contained.
So we observe no large influence of the Yukawa coupling
on the Higgs mass upper bound in our model
for $\kp>0$.


\section{Summary and conclusions}

We have explored the region of the largest renormalized Yukawa coupling
in a lattice Yukawa model with naive fermions
in the broken symmetry phase.
In this region the Yukawa interaction is the driving force of the
spontaneous symmetry breaking
overwhelming the very weak ferromagnetic
or even antiferromagnetic nearest neighbour
scalar field coupling.
Such competing interactions, together with sizeable
fermion loop corrections, result in a complex structure
of the scalar propagators.
We have demonstrated that this structure can be theoretically
understood and even utilized for an extraction of the
renormalized Yukawa coupling $y_R$ from the scalar propagator data.
The results for $y_R$ showing that the Yukawa coupling is
small in the limit of large cutoff are consistent with the triviality
of this coupling.
In the physically relevant FM(W) phase
the lines of constant $y_R$ seem to
flow nearly parallel to the FM(W)-PMW phase transition
and for $y > 1$ run out
of the FM(W) phase, instead of
flowing into some point on this phase transition.
In particular the suspicious point A does not  seem
to be a nontrivial fixed point.
The values of $y_R$ do not seem to exceed significantly
those at $\kp \geq 0$, indicating that the $\kp < 0$ region
of the FM(W) phase does not add much to the
physical content of the model at $\kp \geq 0$.
We expect these results to be generic for various lattice
Yukawa models.

Our results, looked at quantitatively, may however be
specific to the chosen model with a large number
of fermions (32 doublets).
The renormalized Yukawa coupling stays below the tree level
unitarity bound, thus being never strong.
Correspondingly,
no influence of the Yukawa interaction on the
upper bound for the $\sg$-mass could be detected.
However, a word of caution is warranted because
we have not localized reliably the edge of the scaling region,
where the largest values of the renormalized couplings should
actually be determined.
Furthermore, the model suffers from a drawback
caused by the large number of fermions:
the fermion mass generated by the Yukawa
interaction stays substantially lower than the $\sg$ boson mass.
This
makes investigations in the scaling region difficult
with two very different
correlation lengths and
is certainly not a generic feature of Yukawa models.
There are now suggestions of Yukawa models with a small
number (2 and 4) of fermion doublets \cc{Jan92}.
The methods developed in this article should be
appropriate also in these models.
Apart from quantitative differences, it would be
interesting to see if any of the qualitative conclusions
drawn in this paper change.

\vspace{2cm}
{\bf Acknowledgement.}
We thank J.~Smit, M.~Tsypin and F.~Zimmermann for
valuable suggestions and H.A.~Kastrup for discussions
and continuous support.
We have also benefited from discussions with
K.~Jansen, M.~Lindner, I.~Montvay, G.~M\"unster,
J.~Shigemitsu and J.~Vink.
The numerical computations have been performed on the
Cray Y-MP/832 of HLRZ J\"ulich and the S-400 of
RWTH~Aachen.

\vspace{2cm}



\renewcommand{\theequation}{A.\arabic{equation}}
\setcounter{equation}{0}
\noindent{\large{\bf Appendix A:  The metamagnetic Ising model}}

Integrating out the fermionic fields leads to a scalar model with
nonlocal interaction terms.
In this appendix we want to discuss
a simple type of spin model with a
nearest-neighbour ($nn$)- and a next-to-nearest-neighbour
($nnn$)-interaction terms, which appears in the literature in the
context of metamagnets\footnote{See footnote on p.~60 in \cc{KiCo75}.}.
In this model an overlap of the scaling regions
associated with the normal and the staggered magnetizations
occurs analogous to the vicinity of the points A and B
in Yukawa models.

The model is defined by the action
\begin{equation}
S=-2\kappa_{nn}  \sum_{x,\mu>0} \sigma_x \sigma_{x+\mu}
  -2\kappa_{nnn} \frac{8}{24}
  \sum_{x,\mu>\nu} \sigma_x \sigma_{x+\mu+\nu},
\end{equation}
where $\sigma_x$ are Ising spins.

In a mean field approximation the critical lines
are given by \cc{KiCo75}
\bea
\kappa_{nn}^c + \kappa_{nnn}^c = \kp^c_{\rm Ising} \;\;
  &\kappa_{nn}^c>0,\kappa_{nnn}^c>0 &
\mbox{\hspace{1cm} for the FM-PM transition line}  \nonumber \\
-\kappa_{nn}^c + \kappa_{nnn}^c =  \kp^c_{\rm Ising} \;\;
  &\kappa_{nn}^c<0,\kappa_{nnn}^c>0 &                  \nonumber
\mbox{\hspace{1cm} for the AM-PM transition line,}
\eea
where the constant $\kp^c_{\rm Ising}$
is known from the 4-dimensional Ising model to be
about 0.0748. At $\kappa_{nn}^c=0$
the transition lines meet in a multicritical point.
The phase diagram found in a numerical Monte Carlo simulation of the
4-dimensional model shown in fig.~12
has three phases: a ferromagnetic
phase (FM) with $\lag\sigma\rag > 0 ,\lag\sigma^{st}\rag = 0 $,
a paramagnetic phase (PM) with $\lag\sigma\rag = 0 ,
\lag\sigma^{st}\rag = 0 $
and an antiferromagnetic phase (AM) with
$\lag\sigma\rag = 0 ,\lag\sigma^{st}\rag > 0 $.
The phase transition lines separating the symmetric from the
broken phases is of second order and the phase transition separating
FM and AM is of first order.
Except for the absence of the FI phase, this phase diagram is similar
to the phase diagram of the Yukawa model at weak Yukawa coupling,
the triple point being analogous to the point A in fig.~1.

In the vicinity of the multicritical point we find by numerical
simulation that
the inverse propagator of the scalar field in momentum space
has two poles, one of them
at $ap=(\pi, \pi, \pi, \pi)$
and some structure for intermediate momenta.
An example in the ferromagnetic phase near the multicritical point
is shown in fig.~13.
This shape is well described by
the inverse
propagator of a gaussian model with the same ($nn$)- and
($nnn$) kinetic terms
\begin{equation}
S_p=-a_{nn}  \sum_{\mu>0} \{ (1-\cos k_{\mu} \}
  -a_{nnn} \frac{8}{24} \sum_{\mu>\nu}
   \{ 2-\cos(k_{\mu}+k_{\nu})-\cos(k_{\mu}-k_{\nu}) \}  + b  .
\label{eq:Sp}
\end{equation}
The fit of the Monte-Carlo data with the function (\eq{eq:Sp}) plotted
in fig.~13 demonstrates that one can  describe the data
very well in spite of
a nonlinear dependence on $\hw^2$.
If one includes further
interaction terms, the structure of the propagator
at intermediate values of $\hw$ changes.

Of course, integrating
out the fermion field in the Yukawa model produces a scalar theory
which has an infinite number of nonlocal interaction terms.
Already taking only into account the first two orders of
the $1/y$ expansion of the fermion determinant
(see for example \cc{AbShr91}) leads to a large number of
non-single-site terms and also to terms which are no longer
bilinear in the scalar fields.
A small $y$ expansion leads to infinite range interaction terms.
The Yukawa model is therefore much more complicated than the
metamagnetic Ising model
and so its scalar propagators are analysed
in this paper in a different way.
In particular, a gaussian model does not describe the data.
But the qualitative feature
of the propagators, namely the existence of a second pole
at $ap=(\pi, \pi, \pi, \pi)$
if two scaling regions overlap,
can be understood considering simple gaussian and Ising models.



\newpage
\noindent{\large {\bf Figure Captions}}\\

\noindent {\bf Fig.~1}
Phase diagram of the SU(2)$_L \otimes$ SU(2)$_R$
Yukawa model with naive fermions \cc{BoDe90a}.

\noindent {\bf Fig.~2}
Schematic description of the spectrum of scalar states
in the phase diagram around point A.
The staggered quantities are denoted by
the index ``st''. We set $a\!=\!1$ in the figure.

\noindent {\bf Fig.~3}
Schematic possible evolution of the bare Yukawa coupling
constant $y_B$
(its definition is described in the text) along a line of constant
physics $y_R$=const in the FM(W) phase. For comparison also $y_0$
is shown.

\noindent {\bf Fig.~4}
Typical forms of the scalar propagators in the FM phase, mostly
at negative $\kp$.
Plotted is the inverse $\pi$ propagator in the momentum space
as a function of $\hw^2=2 \sum_{\mu} (1-\cos(ap_{\mu}))$.
The left column shows results near the FM(W)--PMW phase transition
and the right column near the FM(S)--PMS transition.
With $\kp$ decreasing downwards the
points A and B, respectively, are approached.
In this figure and in the following, error bars are not shown
whenever they are smaller than the symbols.

\noindent {\bf Fig.~5}
An example of the inverse $\pi$ propagator in the FM(S) phase
close to the point B. The full line is the fit with the
two pole Ansatz analogue to (\eq{stan}).

\noindent {\bf Fig.~6}
An example of the inverse $\pi$ propagator in the FM(W) phase
at small positive $\kp$.
We show in the upper part the Monte Carlo data and
in the lower part the fit with the Ansatz (\eq{ANSATZ}).

\noindent {\bf Fig.~7}
An example of the inverse $\pi$ propagator in the FM(W) phase
at negative $\kp$ close to the point A.
We show in the upper part the Monte Carlo data and
in the lower part the fit with the Ansatz (\eq{ANSATZ})
where a staggered scalar pole is also taken into account.

\noindent {\bf Fig.~8}
Examples of lattice size dependence of $am_F$ and $av_R$.
The dotted and dashed lines are fits linear in $L^{-2}$.
Extrapolations to infinite volume are also shown.

\noindent {\bf Fig.~9}
The wave function renormalization constant $Z_{\pi}$
close to the FM(W)--PMW phase transition
as a function of decreasing $\kp$. The point A is
situated at $\kp=-0.75(3)$.
The dashed line serves to guide the eye.

\noindent {\bf Fig.~10}
Our present results for $y_R$, as defined by the relation
(\eq{TR}), obtained from the infinite volume extrapolated values
of $am_F$ and $av_R$.
The range of the earlier results \cc{BoDe91a} obtained with
a naive analysis of the Goldstone propagators and on
finite lattices in indicated by the dotted error bars
around the dotted curve (\ref{dotc}).

\noindent {\bf Fig.~11}
Ratio $m_{\sigma}/v_R$ as a function of the scalar correlation
length $\xi_{\sigma}$.
The full squares represent a tentative infinite volume extrapolation.
The circles are Monte Carlo results in the pure O(4) $\Phi^4$ theory
\cc{O4uppb}.

\noindent {\bf Fig.~12}
Phase diagram of an Ising metamagnet.
The numerically determined positions of critical lines are
indicated by the points with error bars and the
dotted lines are mean field results.

\noindent {\bf Fig.~13}
Inverse momentum space propagator
in the Ising metamagnet
and its fit by means of
the gaussian model with a $nnn$-term.
\end{document}